\documentclass[aps,prb,preprint,superscriptaddress]{revtex4}

\usepackage{graphicx}
\usepackage{amsmath}
\usepackage{subfigure}
\usepackage[german,english]{babel}
\begin{document}

\title{Theory of optically forbidden d-d transitions in strongly correlated crystals}

\author{M I Katsnelson}
\affiliation{Radboud University of Nijmegen, Institute for
Molecules and Materials, Heijendaalseweg 135, 6525 AJ Nijmegen,
The Netherlands}
\author{A I Lichtenstein}
\affiliation{I. Institut f{\"u}r Theoretische Physik,
Universit{\"a}t Hamburg, Jungiusstra{\ss}e 9, D-20355 Hamburg,
Germany}

\pacs{78.20.Bh; 71.28.+d; 71.10.Fd}


\date{\today}

\begin{abstract}
A general multiband formulation of linear and non-linear optical
response functions for realistic models of correlated crystals is
presented. Dipole forbidden d-d optical transitions originate from
the vertex functions, which we consider assuming locality of
irreducible four-leg vertex.  The unified formulation for second- and
third-order response functions in terms of the three-leg vertex is
suitable for practical calculations in solids. We illustrate the
general approach by consideration of intraatomic spin-flip
contributions, with the energy of 2$J$, where $J$ is a Hund exchange,
in the simplest two-orbital model.
\end{abstract}
\maketitle


From the physical point of view the most of natural mineral dyes are
the Mott or charge-transfer insulators and their colors are
determined essentially by correlation effects\cite{optics}. If a material has no
energy gap or its value is smaller than the energy of visual light, it
will be non-transparent, a black one or with metallic shine. In the
case of  broad-gap materials the absorbtion of visual light is
determined by impurities (ruby, that is, Al$_{2}$O$_{3}$ doped with
Cr$^{3+}$ ions is a prototype example \cite{optics}) or by optically
(dipole) forbidden d-d transitions between different terms and
multiplets belonging to the same d$^{n}$-configurations of
transition metal ions for a pure system. The latter processes are
responsible for a green color of NiO\cite{NiO} and blue color of
most of the divalent copper insulating compounds\cite{Pisarev}.
These systems are usually colored and transparent and the
transparency itself is a manifestation of dipole-forbidden character
of relevant optical transitions.

Up to now the optical properties of Mott or charge transfer
insulators are considered within the framework of cluster approaches\cite{cluster,millis}. The
present paper develops a general translationally invariant formalism
to treat the d-d transitions in strongly correlated
$\mathit{crystals}$. It is commonly accepted now that the standard
LDA(GGA) approach is insufficient to describe the electronic
structure of the Mott insulators\cite{Kubler}. To have more adequate
picture of single-electron spectra various approaches have been
applied to the problem such as LDA+U\cite{Anisimov},
self-interaction corrections\cite{Svane-Gunnarsson} and the
GW-scheme\cite{Schilfgaarde}. However, all these approaches do not
provide the correct atomic limit and in particular do not take into
account the term and multiplet structure, which is crucial for
optics. This problem can be solved within the LDA+DMFT (dynamical
mean-field theory)\cite{LDAplusDMFT,LDAplusplus} and the Hubbard-I
approximation\cite{LDAplusplus,SvaneH1}. There were several
attempts to calculate the optical properties within the LDA+DMFT
using the Kubo formula for optical
conductivity\cite{Savrasov-Udovenko,Chadov,Georges-optic}. However,
in all these calculations the vertex contributions were not taken
into account and the two-particle Green functions were calculated as
a convolution of two single-particle Green functions (for review of
the LDA+DMFT see\cite{KotliarLDA+DMFT}). The latter contains only
transitions related to the promotion of d-electrons to the p-band
with the change of transition-metal configurations from d$^{n}$ to
d$^{n \pm 1}$ and thus this approach is not sufficient to explain
why NiO is green.

In the DMFT approach the self-energy is local which leads to a
cancellation of vertex corrections in the single-band Hubbard
model\cite{George-DMFT}. However, this is not the case for a generic
multiband situation, similar to a treatment of optical properties of
disordered alloys in the coherent-potential approximation (CPA)
\cite{Buttler}. Here we present the corresponding formulation for
linear and non-linear optical response functions.


We start with the general expression for linear optical conductivity
\begin{equation}
\sigma _{ab}(i\omega )=\frac{e^{2}}{\omega }T^{2}\sum_{\nu \nu ^{\prime
}}\sum_{1234}\left\langle 4\left| v_{a}\right| 1\right\rangle \left\langle
2\left| v_{b}\right| 3\right\rangle \chi _{1234}(i\nu ,i\nu ^{\prime
},i\omega )  \label{sigma}
\end{equation}
where $\chi _{1234}$ is a generalized susceptibility in a
quasiatomic basis set $\left| 1\right\rangle =\left| iL\sigma
\right\rangle $, where $i,L,\sigma
$ label sites, orbital quantum numbers and spin projections, respectively, $%
v_{a}$ is the electron velocity operator ($a=x,y,z$), and we use the
Matsubara
Green functions; at the end of calculations the analytical continuation $%
i\omega \rightarrow \omega +i0$ should be performed\cite{Mahan}.

The two-particle Green function $\widehat{\chi }$ is expressed in
terms of the single-particle Green function $\widehat{G}$ and the
irreducible vertex function $\widehat{\Gamma }$ as
\cite{George-DMFT,Migdal}:
\begin{eqnarray}
\chi _{1234}(i\nu ,i\nu ^{\prime },i\omega ) &=&-G_{12}(i\nu ^{\prime
})G_{34}(i\nu )\delta _{\nu ,\nu ^{\prime }+\omega }- \\
&&-T\sum_{\nu ^{\prime \prime }}\sum_{5678}G_{15}(i\nu ^{\prime
})G_{84}(i\nu )\Gamma _{5678}(i\nu ,i\nu ^{\prime \prime },i\omega )\chi
_{6237}(i\nu ^{\prime \prime },i\nu ^{\prime },i\omega )
\end{eqnarray}
which can be written in the matrix form in the fermionic Matsubara
frequencies $(i\nu ,i\nu ^{\prime })$ and the pairs of electron
quantum numbers $(14,23)$ as
\begin{equation}
\hat{\chi} =\hat{\chi}_{0}+\hat{\chi}_{0}\hat{\Gamma}\hat{\chi}   \label{xi}
\end{equation}
where $\hat{\chi}_{0}=-\hat{G}\ast \hat{G}$. Within the DMFT approximation
the self-energy $\widehat{\Sigma }(i\omega )$ is local, that
is, diagonal in site indices and $\mathbf{k}$-independent in the
momentum representation\cite {George-DMFT}. In addition we will
assume a locality of the irreducible vertex function
$\widehat{\Gamma }$; this is the only approximation we add. Than,
it can be obtained from the local version of Eq.(\ref{xi}):
\begin{equation}
\hat{\Gamma} =\hat{\chi}_{loc,0}^{-1}-\hat{\chi}_{loc}^{-1}  \label{gamma}
\end{equation}
where $\hat{\chi}_{loc,0}$, $\hat{\chi}_{loc}$ are matrices in the Matsubara
frequencies and pairs of orbital and spin indices, all site indices
are supposed to be the same\cite{George-DMFT}. Both single-particle
and two-particle \textit{on-site} Green functions can be found
numerically exactly using full diagonalization
scheme\cite{Hafermann} or continuous-time Quantum Monte
Carlo method\cite{Rubtsov}.

\begin{figure}[t]
\begin{center}
\includegraphics[width=0.3\linewidth]{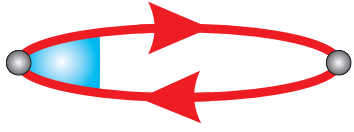}
\end{center}
\caption{\label{fig:cond_Kubo}(Color online) Diagrammatic
representation of Eq.(\ref{sigma1}); thick lines are the exact Green
functions, dot is the bare matrix element $\left\langle
1\mathbf{k}\left| v_{a}\right| 2\mathbf{k}\right\rangle $ and dot
with triangle corresponds to $\left\langle 1\mathbf{k}\left|
v_{a}^{eff}(i\nu ,i\omega )\right| 2\mathbf{k}\right\rangle $.
\label{Fig1}}
\end{figure}

Thus, the optical conductivity (\ref{sigma}) can be expressed in the
following form:
\begin{equation}
\sigma _{ab}(i\omega )=-\frac{e^{2}}{\omega }T\sum_{\nu }\sum_{\mathbf{k}%
}\sum_{1234}\left\langle 4\mathbf{k}\left| v_{a}^{eff}(i\nu ,i\omega
)\right| 1\mathbf{k}\right\rangle G_{12}(\mathbf{k,}i\nu )
\left\langle 2\mathbf{k}\left|
v_{b}\right| 3\mathbf{k}\right\rangle G_{34}(\mathbf{%
k,}i\nu +i\omega )  \label{sigma1}
\end{equation}
where b$1234$) are orbital and spin indices only and the effective
matrix element (three-leg vertex) satisfies the equation
\begin{eqnarray}
\left\langle 4\mathbf{k}\left| v_{a}^{eff}(i\nu ,i\omega )\right| 1\mathbf{k}%
\right\rangle  &=&\left\langle 4\mathbf{k}\left| v_{a}\right| 1\mathbf{k}%
\right\rangle -T\sum_{\nu ^{\prime }}\sum_{\mathbf{k}^{\prime
}}\sum_{2356}\left\langle 3\mathbf{k}^{\prime }\left|
v_{a}^{eff}(i\nu ^{\prime},i\omega )\right| 2\mathbf{k}^{\prime}\right\rangle   \label{g3} \\
&&G_{25}(\mathbf{k}^{\prime}, i\nu ^{\prime})
G_{63}(\mathbf{k}^{\prime}, i\nu ^{\prime}+i\omega)\Gamma
_{5146}(i\nu ^{\prime}, i\nu ,i\omega ) \nonumber
\end{eqnarray}
Diagrammatically Eq.(\ref{sigma1}) is shown in Fig.\ref{Fig1}.

\begin{figure}
\subfigure[]{
\includegraphics[scale=0.3]{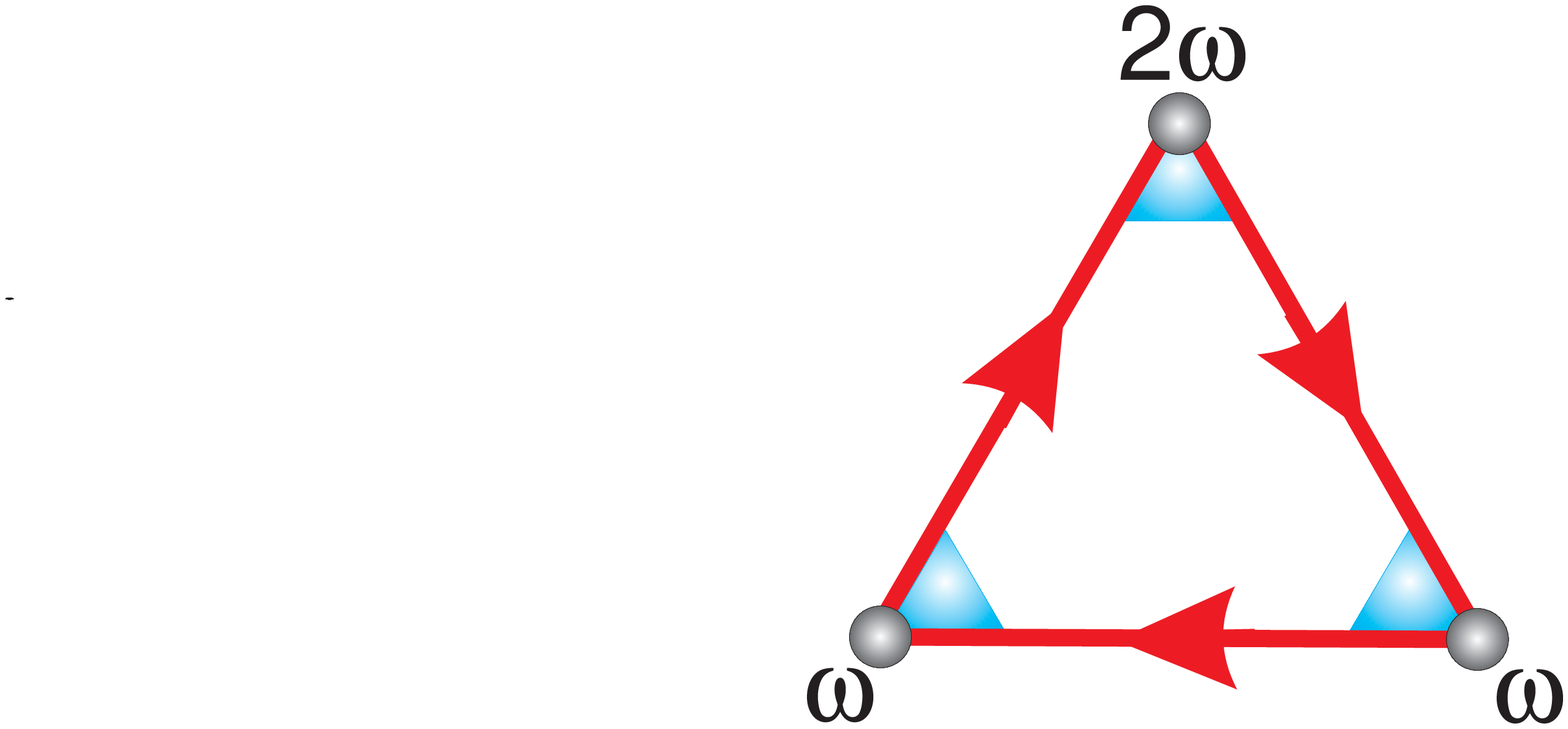}
}
\subfigure[]{
\includegraphics[scale=0.3]{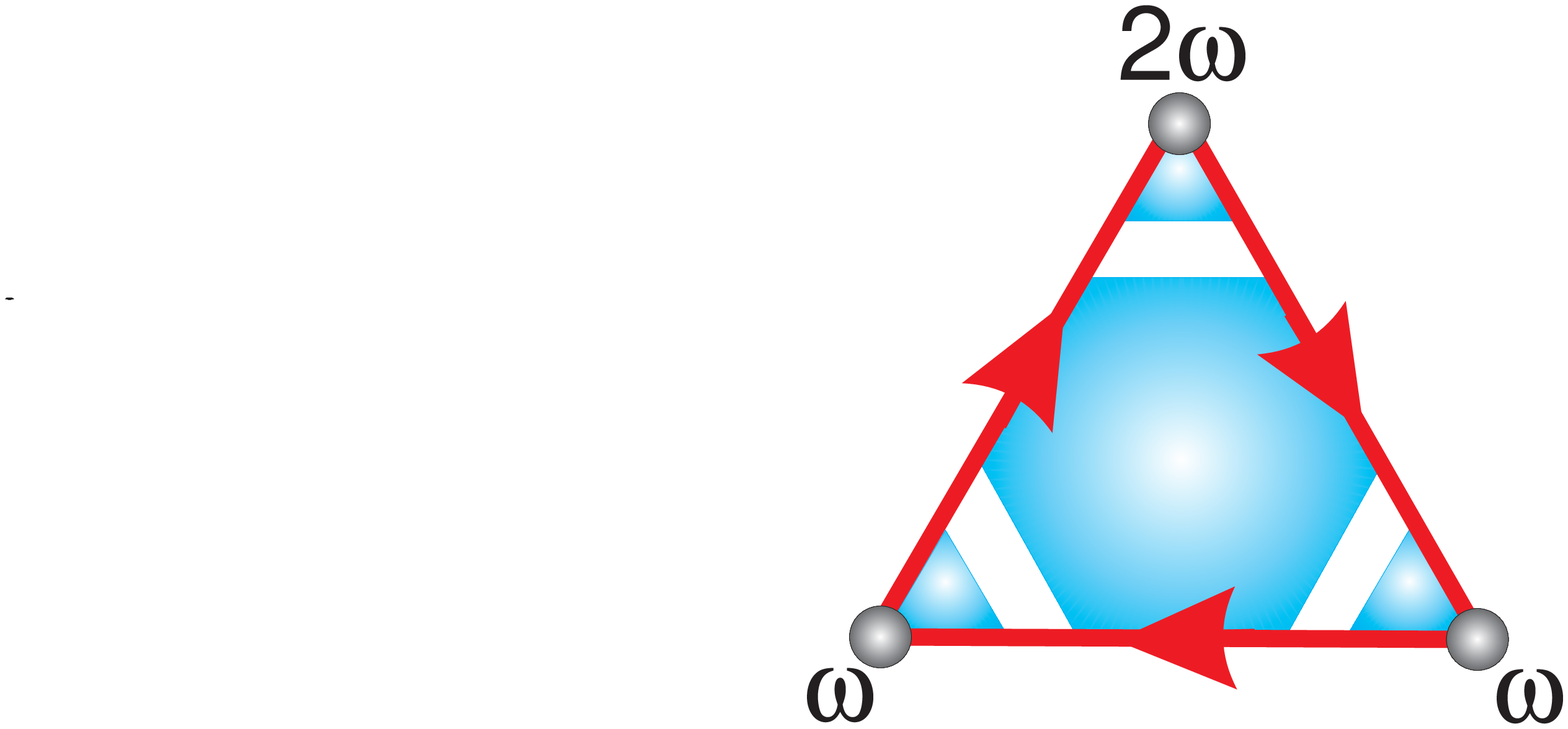}
}
\caption{(Color online) Diagrammatic
representation of the non-linear susceptibility for second harmonic
generation;  thick lines are the
exact Green functions,  dot with triangle corresponds to $\left\langle 1%
\mathbf{k}\left| v_{a}^{eff}(i\nu ,i\omega )\right|
2\mathbf{k}\right\rangle $ and the shadowed hexagon is the
irreducible six-leg vertex. }
\label{Fig2}
\end{figure}

The effective matrix element $\left\langle 1\mathbf{k}\left|
v_{a}^{eff}(i\nu ,i\omega )\right| 2\mathbf{k}\right\rangle $ is
convenient since its use allows us to present the linear and
non-linear response functions in the unified form. For example, the
non-linear optical susceptibility describing a second harmonic
generation can be \textit{exactly} represented as a sum of diagrams
shown in Figs.\ref{Fig2}a and \ref{Fig2}b.

The corresponding analytical expression for Fig.\ref{Fig2}a reads
\begin{eqnarray}
\label{Sigma3}
&&\chi _{abc}(i\omega ,i\omega ,2i\omega ) = \frac{e^{2}}{\omega ^{2}}%
T\sum_{\nu }\sum_{\mathbf{k}}\sum_{123456}\left\langle 6\mathbf{k}\left|
v_{a}^{eff}(i\nu ,i\omega )\right| 1\mathbf{k}\right\rangle
G_{12}(\mathbf{k,}i\nu)
\\
&&
\left\langle 2%
\mathbf{k}\left| v_{b}^{eff}(i\nu +i\omega ,i\omega )\right| 3\mathbf{k}%
\right\rangle
G_{34}(\mathbf{k,}i\nu + i\omega )
\left\langle 4\mathbf{k}\left| v_{c}^{eff}(i\nu - i\omega
,2i\omega )\right| 5\mathbf{k}\right\rangle
G_{56}(\mathbf{k,}%
i\nu -i\omega )  \nonumber
\end{eqnarray}

The calculations of magnetic susceptibility in the one-band Hubbard
model\cite{HafermannPRL}  shows that the contributions of the
six-leg vertex are small; one can hope that this is also the case
for the second harmonic generations and thus Eq.(\ref{Sigma3}) will
be sufficient for real calculations.

We proceed with the multiband Hubbard model with the Hamiltonian
\begin{equation}
H=\sum_{im\sigma }\varepsilon _{i}c_{im\sigma }^{+}c_{im\sigma }+\sum_{\substack{ %
i\neq j \\ mm^{\prime }\sigma}}  t_{mm^{\prime }}^{ij}c_{im\sigma
}^{+}c_{jm^{\prime }\sigma }+\frac{1}{2}\sum_{\substack {i\sigma \sigma ^{\prime } \\ %
m_{1}m_{2}m_{3}m_{4}}}  U_{m_{1}m_{2}m_{3}m_{4}}c_{im_{1}\sigma
}^{+}c_{im_{2}\sigma ^{\prime }}^{+}c_{im_{4}\sigma ^{\prime
}}c_{im_{3}\sigma }  \label{Ham}
\end{equation}

\begin{figure}
\centering
\subfigure[]{
\includegraphics[scale=0.26]{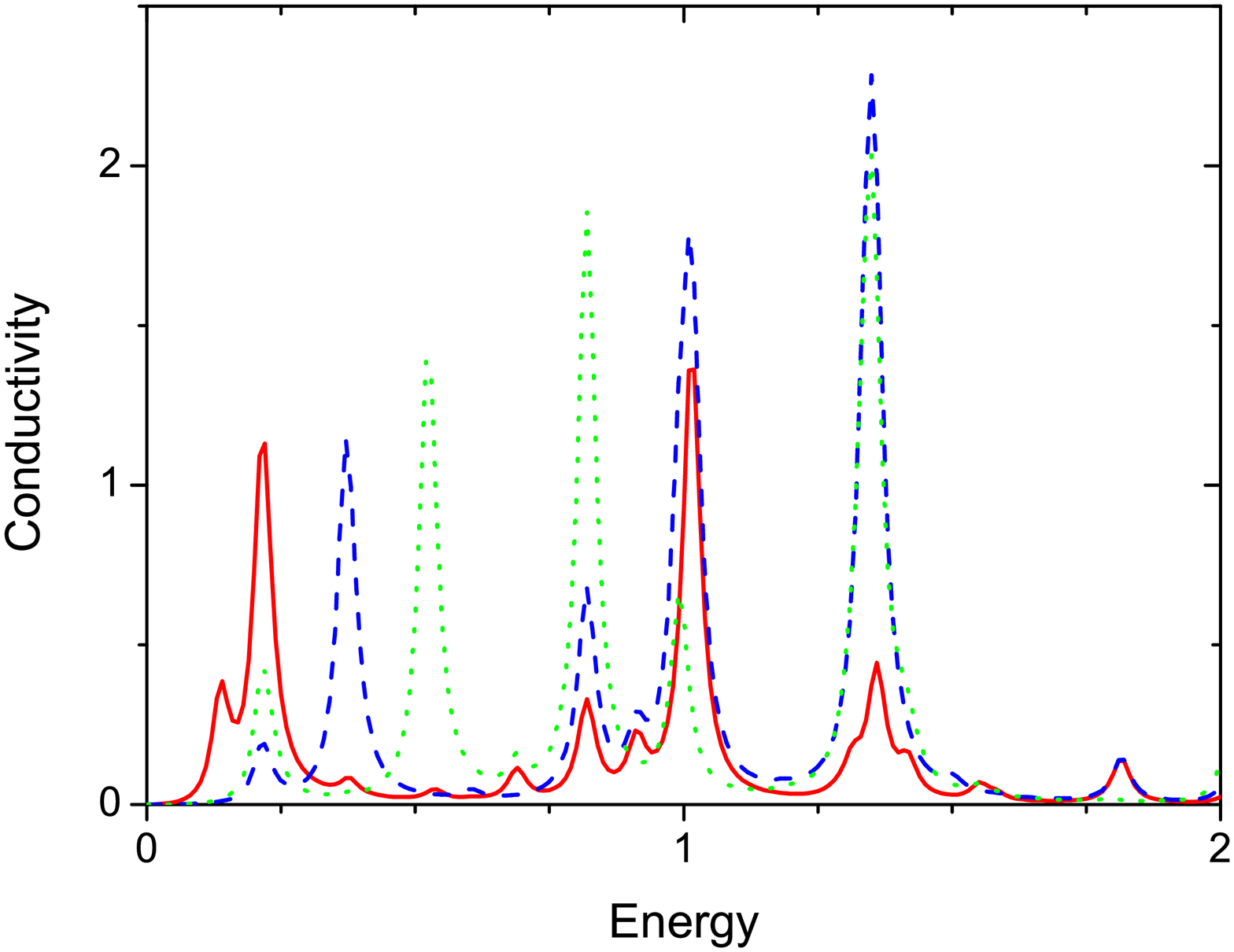}
}
\subfigure[]{
\includegraphics[scale=0.26]{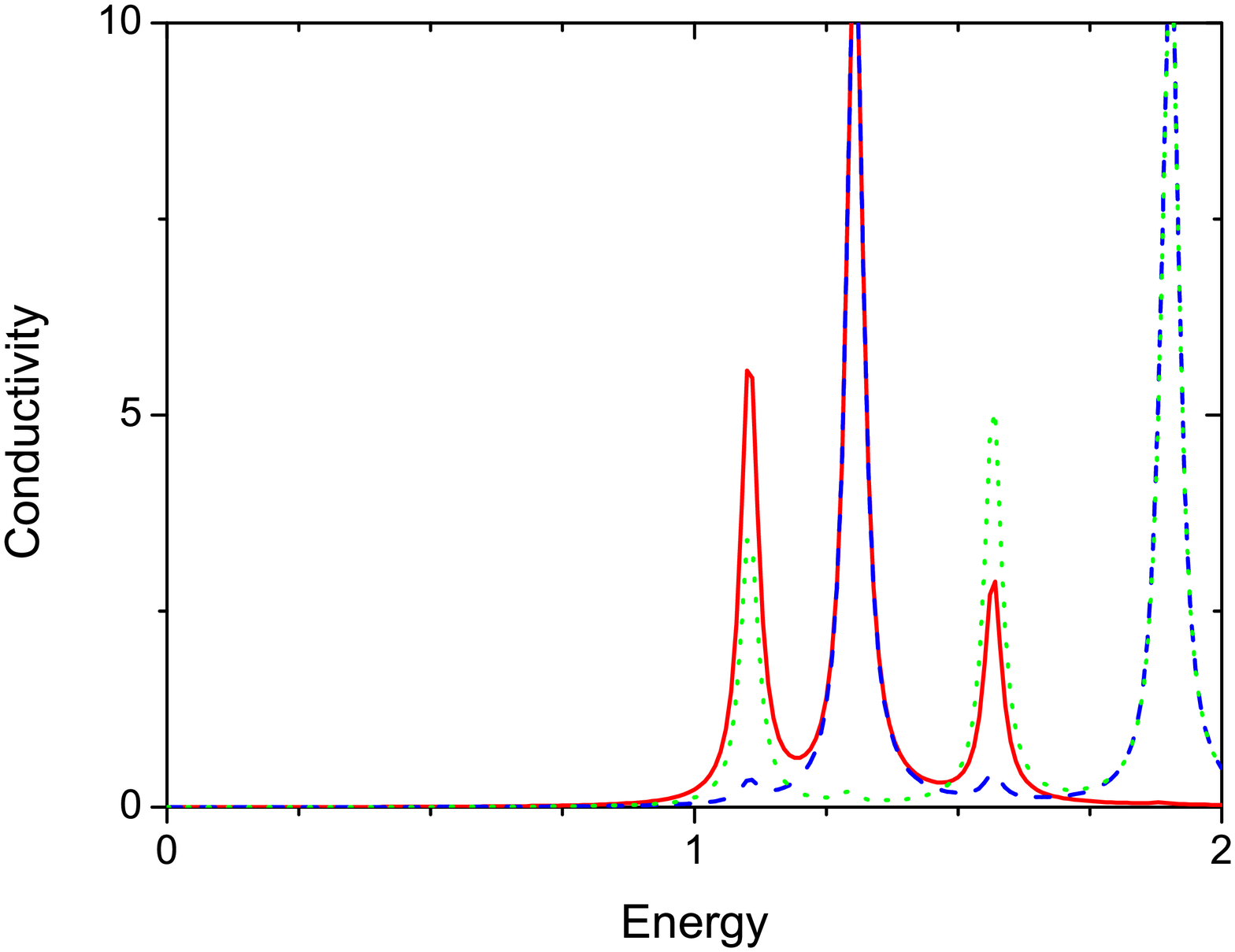}
}
\caption{(Color online) Components of generalized
susceptibility $\rm{Im} \chi _{im,jm^{\prime };im,jm^{\prime }}/\omega$ summed up over ($%
\sigma ,\sigma ^{\prime }$) for two-site two-band model with $\varepsilon
_{1}=\varepsilon _{2}=0$, $U=1$, $J=0.2$, $t_{12}=t_{21}=0.05$; (a): $%
t_{11}=t_{22}=0.2$; (b) $t_{11}=t_{22}=0.5$. Solid red curve
corresponds to intrasite interband transitions $im=(11)$,
$jm^{\prime }=(12)$, dashed blue curve -- to intersite intraband
transitions $im=(11)$, $jm^{\prime }=(21)$, and
the dotted green curve -- to intersite interband $im=(11)$, $jm^{\prime }=(22)$%
. We use the spectral representation with the $\rm{Im}\omega =0.05$
and the temperature is $T=0.05$. Energies are in the units of $U$. \label{Fig3}}
\end{figure}

\begin{figure}
\centering
\subfigure[]{
\includegraphics[scale=0.26]{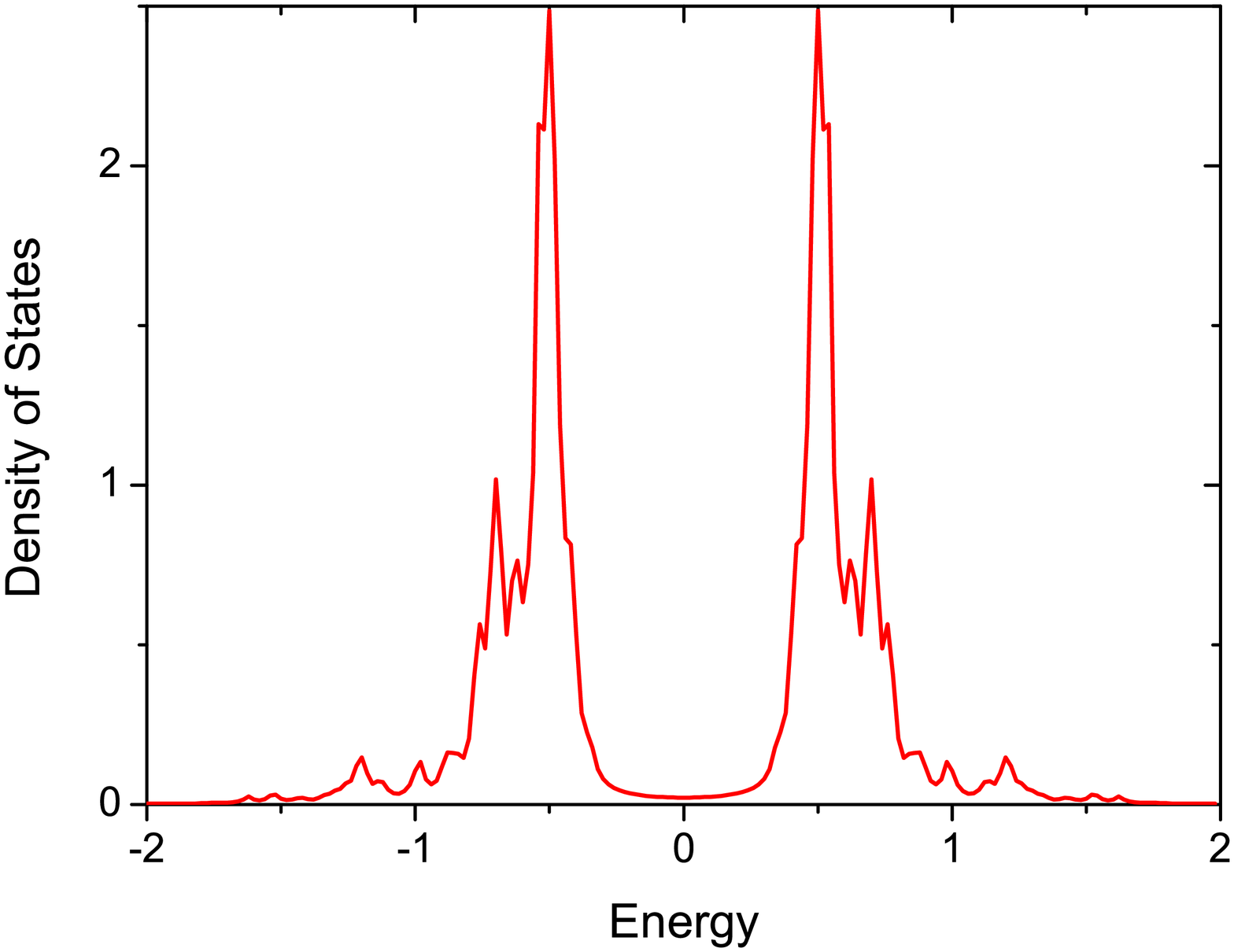}
}
\subfigure[]{
\includegraphics[scale=0.26]{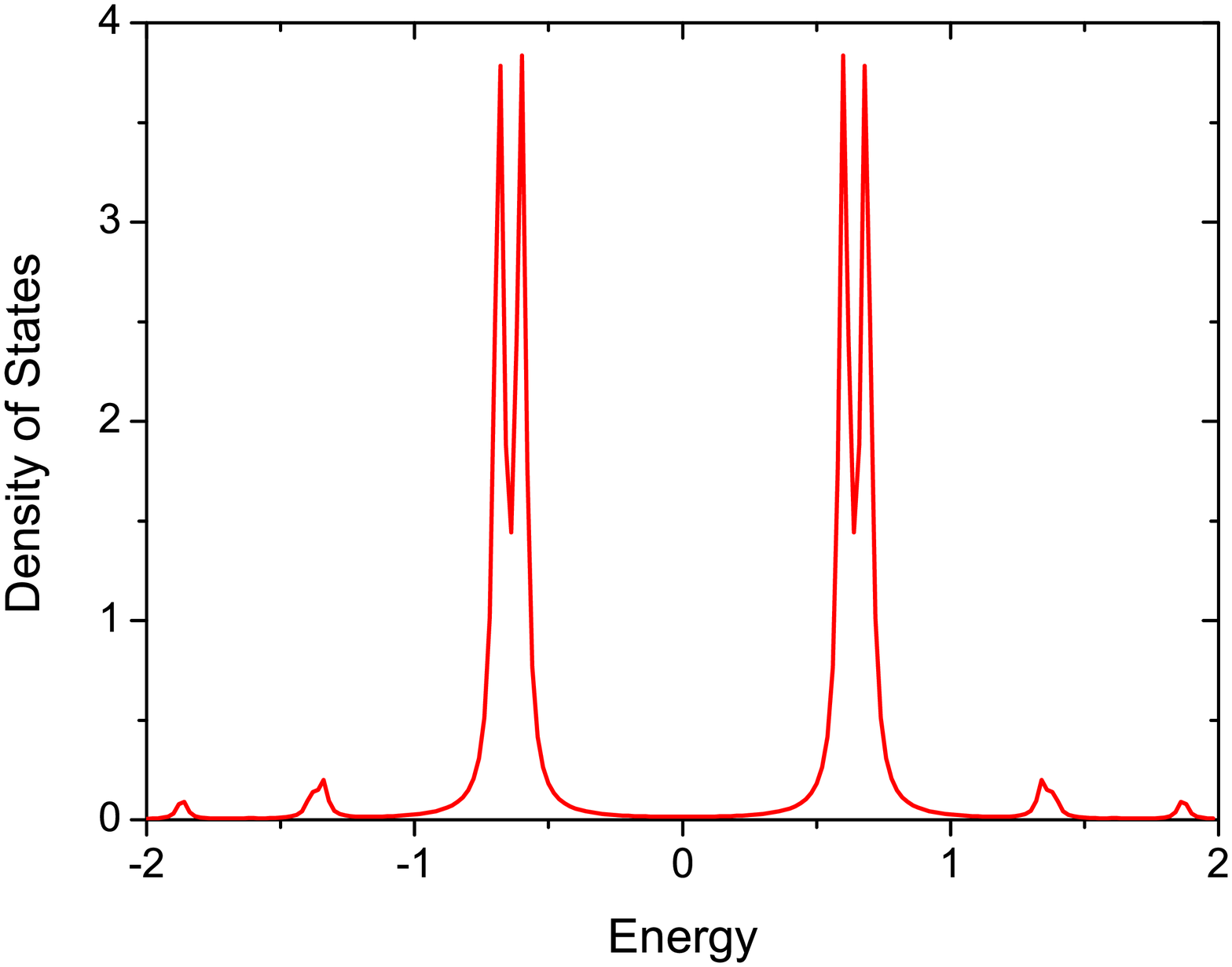}
}
\caption{(Color online)
Density of single-particle
states for the same parameters as in Fig.3 Energies are in the units of $U$.\label{Fig4}}
\end{figure}

To clarify a physical meaning of vertex corrections to the response
functions we discuss first exactly solvable model of two sites with
two orbitals ($i=(1,2)$, $m=(1,2)$). The corresponding rotationally
invariant interaction matrix is parametrized by the Hubbard energy
$U$ and the Hund exchange parameter $J$ as\cite{Fresaar}
\begin{eqnarray}
U_{m_{1}m_{1}m_{1}m_{1}} &=&U  \nonumber \\
U_{m_{1}m_{2}m_{1}m_{2}} &=&U-2J  \nonumber \\
U_{m_{1}m_{2}m_{2}m_{1}} &=&J  \label{U}
\end{eqnarray}
($m_{1}\neq m_{2}$). The dimension of the Hilbert space is equal to $%
2^{8}$ so $\widehat{\chi }$ and $\widehat{G}$ can be easily
found by exact diagonalization. The results for $\rm{Im} \hat{\chi}/
\omega \propto \rm{Re} \hat{\sigma} (\omega)$ for real (not Matsubara)
frequencies are shown in Fig.\ref{Fig3} and Fig.\ref{Fig4}.

As one can see from the density of states the single-particle
transitions which manifest themselves in $\hat{\chi}_{0}$ starts at the
frequency $\omega \geq 1 $ corresponding to the distance between the
highest occupied and the lowest unoccupied orbitals. For small
enough hopping (Fig.\ref{Fig3}a) there is a lot of peaks at smaller
energies which originate from the poles of the vertex functions and
represent the optically forbidden transitions in our toy model. In
particular, a transition with $\omega =2J=0.4$ is clearly
visible for intraband intersite transitions $(11)\rightarrow (21)$. The transition corresponds to local intraatomic spin flip
processes. It is visible in optical (summed up over spins)
susceptibility, since the Coulomb interaction matrix couples spin-up
and spin-down states. An interesting and unexpected result of the
toy model is that for a moderate hopping (Fig.\ref{Fig3}b) all
these local term effects disappear.

Now we present the main part of our work related with the two-band
lattice model with the use of local approximation for vertex $\hat{\Gamma}$ as describes above. The calculations have been done for the square
lattice in the nearest neighbour approximation. The local vertex
function has been obtained from Eq.(\ref{gamma})  by exact
diagonalization calculations of $\widehat{\chi }$ and $\widehat{G}$.
If the exact diagonalization would be performed for intraatomic
Coulomb interaction Hamiltonian only, this would correspond to the
Hubbard I approximation\cite{LDAplusplus}. To go beyond this, in
spirit of the DMFT\cite{George-DMFT}, we have added one more
orbital to the bath.

\begin{figure}[t]
\begin{center}
\includegraphics[width=0.5\linewidth]{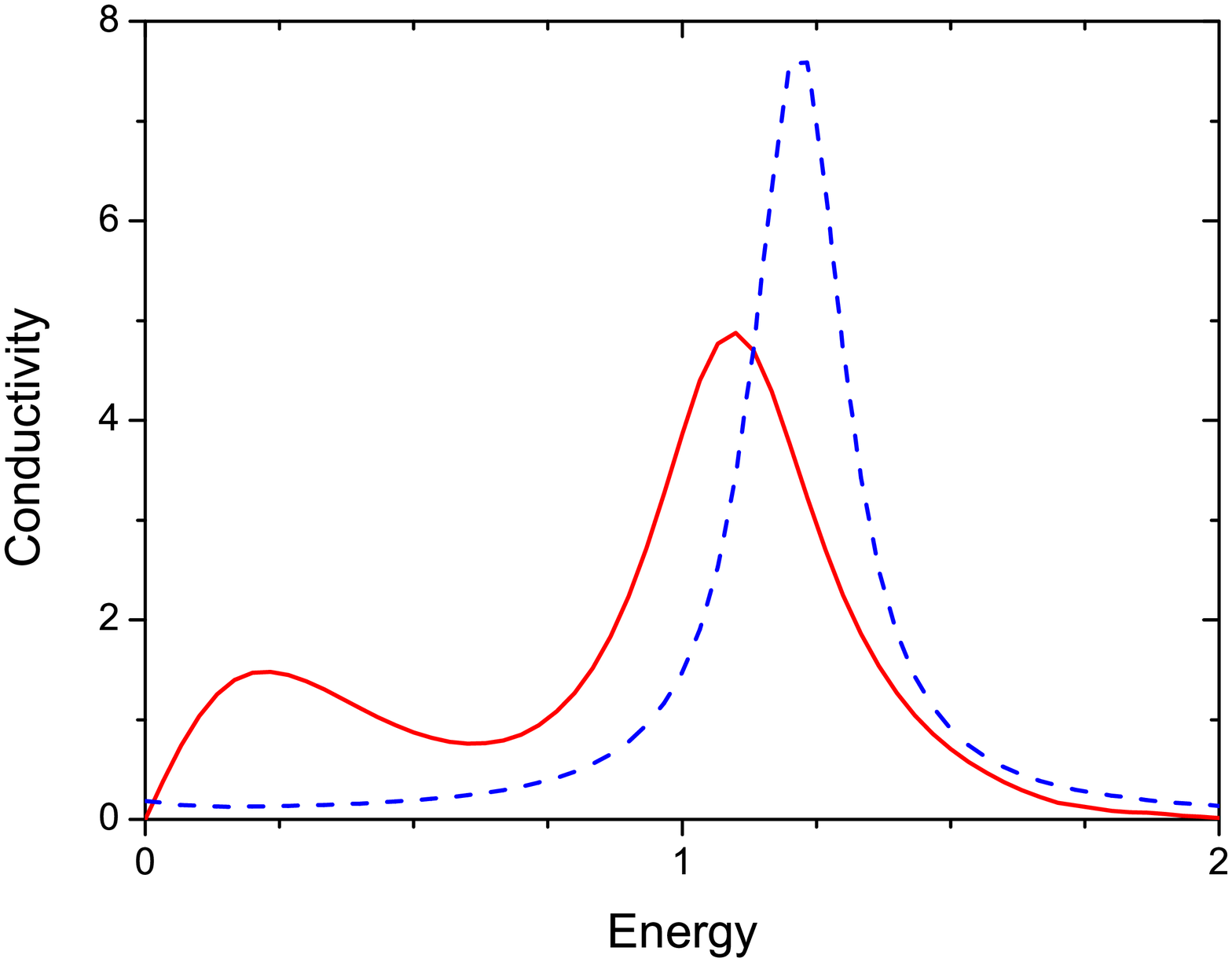}
\end{center}
\caption{(Color online) Intrasite, interband components of
$\rm{Im}\widehat{\chi}/\omega \propto \rm{Re} \sigma (\omega)$
(solid red curve) and $\rm{Im}\widehat{\chi}_0/\omega \propto
\rm{Re} \sigma_0 (\omega)$ (dashed blue curve) for the square
lattice in the nearest-neighbour approximation, with the same
parameters as in Fig.3a. Energies are in the units of $U$.\label{Fig5}}
\end{figure}

Despite simplicity of the model our calculations turned out to be
rather time and memory consuming due to the inversion of
$\widehat{\chi }$-matrices depending of two Matsubara frequencies
$(i\nu ,i\nu ^{\prime })$. To reach convergence, we had to use about
one hundred frequencies and $20\times 20$ $\mathbf{k}$-points. The
computational results are shown in Fig.\ref{Fig5}. One can see that
$\widehat{\chi }_{0}$ has only one pronounced peak at $\omega =U$
corresponding to the transition from the lower to upper Hubbard
bands. At the same time, $\widehat{\chi }$ has an additional peak
within the gap of the single-particle excitation spectrum at the
maximum around $\omega =2J=0.4$. This maximum originates from the
spin-multiplet structure of the d$^{2}$ configuration.

To conclude, we present a general formalism which allows to consider
term and multiplet effects on linear and non-linear optical
properties of multiband strongly correlated systems. For a two-band
model the computational results are quite reasonable and one can
hope that the scheme can be applied to the first-principle
investigations of realistic systems in the spirit of the LDA+DMFT.
The approach can be apply for other response functions such as
magnetic susceptibilities and STM-spectroscopy where the transition
with energy $\omega =2J$ have been recently observed for Mn-chains
on Pt surface\cite{IBM}.

\end{document}